# Toward an Actionable Socioeconomic-Aware HCI

Margaret Burnett, Abrar Fallatah, Catherine Hu, Christopher Perdriau,
Christopher Mendez, Caroline Gao, Anita Sarma

*Oregon State University, Corvallis, Oregon 97331 USA*

Although inequities for individuals in different socioeconomic situations are starting to capture widespread attention, less attention has been given to the socioeconomic inequities that saturate socioeconomic-diverse individuals' *user experiences.* To enable HCI practitioners to attend to such inequities and avoid unwittingly introducing them, in this paper we consider a wide body of research relevant to how an individual's socioeconomic status (SES) can affect their user experiences with technology. We synthesize this foundational research to produce a core set of 6 evidence-based SES "facets" (attribute types and value ranges) that directly relate to user experiences for individuals in different SES strata. We then harness these SES facets to produce actionable paths forward—including a new structured method we call SocioeconomicMag—by which HCI researchers and practitioners can bring new socioeconomic-aware practices into their everyday HCI work.

## 1 INTRODUCTION

Individuals who are experiencing lower socioeconomic status (SES) make up a sizeable fraction of the populations that technology purports to serve. For example, in 2017 in the U.S. alone, 39 million people were living below the poverty line [70]. But socioeconomic status goes beyond its tie to an individual's income. It also impacts their educational, occupational, and social opportunities and privileges—and the attitudes and perceptions they develop from their experiences in light of these impacts. Thus, it can have profound effects on individuals' interactions with technology, and on technology's impacts on individuals, as we show in later sections.

Experts disagree on exactly which factors determine someone's SES, so in this paper we deliberately do not define it. Although most researchers agree that the factors of income, education, and occupation are part of the definition [205, 130], most researchers also point out that these three factors alone are insufficient [197]. As Braveman et al. put it, "one size does not fit all" [28]; i.e., researchers must consider the population being studied to choose an applicable definition for that population. Braveman et al. suggest that researchers need to measure as many relevant socioeconomic factors as possible, including race, ethnicity, social class, food security and as many others as possible that are pertinent to the population being studied [20, 28, 94]. In this paper we accept the particular SES definition used in each work from which we draw.

On the surface, it might seem that socioeconomic inequity is not an HCI problem, but in this paper we will "build on the shoulders of giants" whose work comes together to show that it is. Still, a gap in this body of work is that, although prior researchers have already investigated selected low-SES subpopulations' technology needs and successes (enumerated in Sections 2–3), each of these works is a distinct point in the research space. What that space currently lacks is a crosscutting view of these works that would enable the HCI community to "connect the dots" in actionable ways. Toward that end, we present a synthesis of these works with an eye toward actionability. From this synthesis, we then derive actionable ways by which today's HCI practitioners can remove some of the SES-inclusiveness barriers that past generations of technology creators have unwittingly erected.

Thus, the contributions of this paper are:

- *Foundations for a Socioeconomically-Aware HCI:* A synthesis of over 200 research papers from multiple fields that can contribute to a more socioeconomically-aware HCI;
- *SES Facets*: Six evidence-based SES "facets" (attribute types and value ranges) derived from these foundations, that directly relate to SES-diverse individuals' user experiences;
- *Insights*: A set of 12 evidence-based insights derived from the foundations.
- *Mapping between Facets and Design Recommendations*: A many-to-many mapping between the SES facets and design recommendations; and

- Multiple *Actionable Steps* to enable HCI researchers and practitioners to add socioeconomic awareness to their HCI practices, including a new systematic inclusive design method we call *SocioeconomicMag*.

*Researcher Self-Disclosure*: Our research team members identify as multiple races (Asian, Black, Latinx, White), with national/ethnic backgrounds from Africa, Asia, the Middle East, and North America. We are all SES-secure at the present time, but some of us have spent substantial portions of our lives as lower-SES individuals. We recognize that, as academic researchers, we are in positions of privilege. We are committed to using that privilege to contribute to technology's inclusive design, to help avert user experience design decisions that disproportionately disadvantage groups of people at any socioeconomic level.

## 2 THE CANDIDATE FACETS

In Section 1 we defined facets as attribute types with associated value ranges. Specifically for SES, we now define an *SES facet* as an attribute type whose values have correlations with individuals' SES, and an *SES facet value* as the particular value some individual has for that facet. For example, if a (hypothetical) SES facet were "highest level of education achieved", one individual's facet value might be "some college". A central design goal was for the SES facets to have values near one end of an SES facet's range that statistically cluster around low-SES individuals (e.g., some high-school), and values near the other end of that facet's range that statistically cluster around high-SES individuals (e.g., graduate degree).

In addition, we impose the following constraint: each facet must have a *wide range* of values, for two reasons. First, a wide range can describe a broad spectrum of individuals' values for that facet. Second, if the facets encourage a designer to create designs that simultaneously support both ends of a facet range, then the wider the range of values, the more facet values *within* that range that will be supported, and therefore the greater the number of individuals that can be inclusively supported by that design.

To find SES facet candidates, we began by surveying scholarly papers (e.g., journals, papers, books, and theses) related to the impact of socioeconomic status on people's technology experiences. Along the way, we also got feedback from experts in aspects of SES, and added literature to our survey as per their recommendations. Finally, we added statistical and census data from government and other public sources. Thus, this paper is not a systematic literature review or a systematic mapping paper: we did not restrict our exploration of the literature to using only specific search strings, specific databases, etc., and do not quantitatively summarize the area. This paper is instead a synthesis of foundational literature's implications for SES-awareness that HCI practitioners can use.

The papers we surveyed ranged from foundational sociological research to empirical studies on how particular populations with particular SES attributes used various technologies. We conducted this research over a period of about three years, ultimately surveying over 200 papers.

As a result, we identified 20 facet candidates. Our inclusion criteria for a facet candidate were:

- At least one scholarly paper provided evidence of the facet's potential importance to people in at least one SES stratum; and
- The facet candidate had a wide range of possible values tied to different SES strata; and
- The facet was potentially relevant to the way people in different SES strata might use technology.

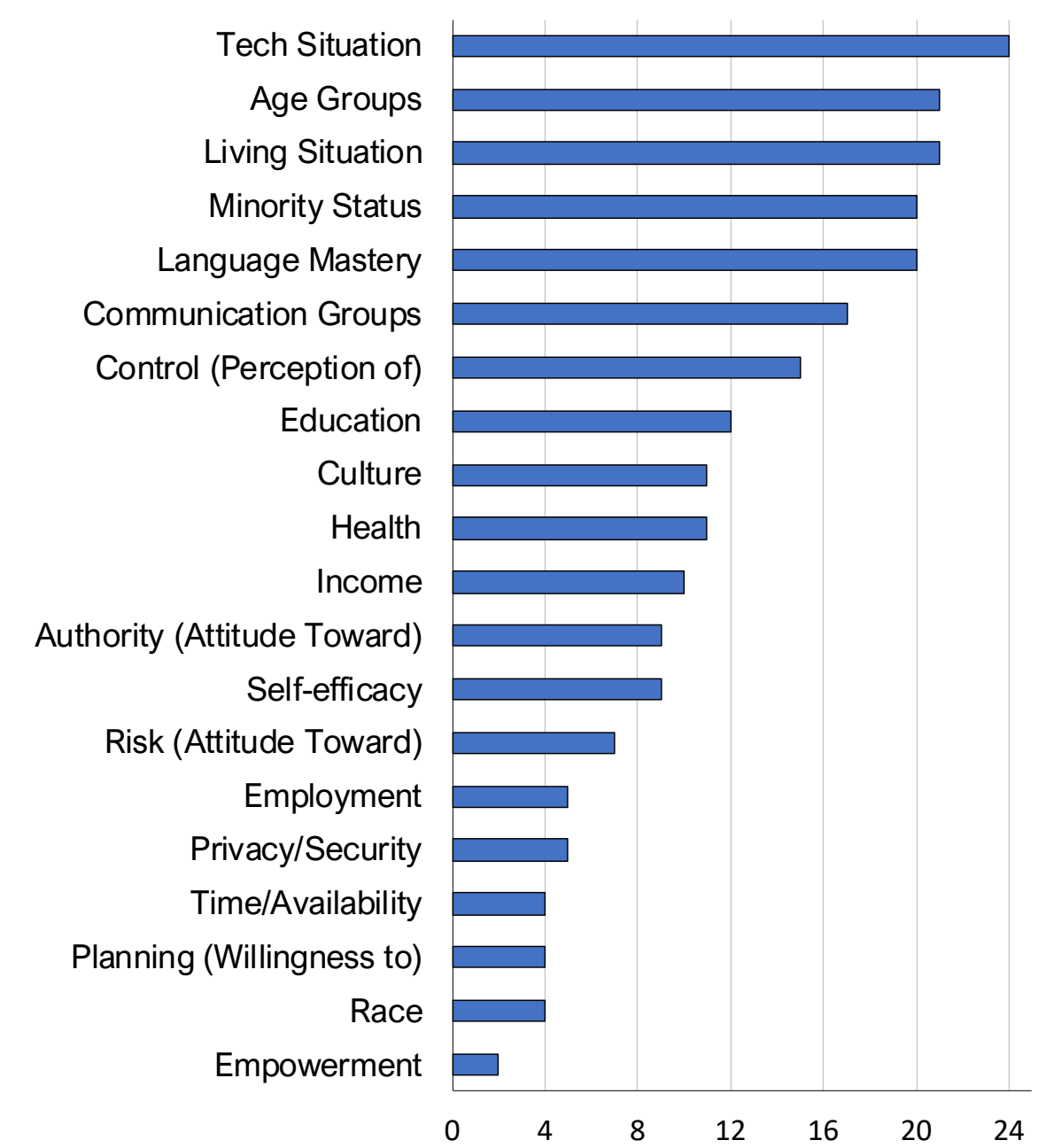

Figure 1: Number of papers that produced each facet candidate detailed in Appendix A.

The 20 facet candidates meeting these criteria are detailed in Appendix A and summarized in Figure 1.

## 3 SIX SOCIOECONOMIC FACETS

Our aim was an *actionable* set of facets for HCI practitioners, so we applied the following inclusion criteria to Appendix A's facet candidates, which resulted in a set of six final facets and facet values. Criteria 1–4 are for each *individual* facet:

- Criterion 1: The facet candidate is one of, or a combination of, the facet candidates from Appendix A;
- Criterion 2: At least five independent studies or papers provide evidence of the facet's ties to SES characteristics, showing disproportionate clusters of lower- and higher-SES individuals at different ends of the facet's spectrum of values;
- Criterion 3: The facet is understandable and usable by software practitioners with no background in social sciences, such as sociology, psychology, race/ethnic studies, etc.;
- Criterion 4: The facet has implications for technology that are clearly applicable to technology design decisions that could be made.

In addition, we required the final set of facets to satisfy the following criteria as a *set*:

- Criterion 5: A small enough set so that they can be kept in mind all at once;
- Criterion 6: Together, the set can capture a large enough subset of the diverse characteristics of the population, that designing for the entire range of values for each facet can make an impactful improvement in the socioeconomic inclusivity of the technology.

Many of Appendix A's facet candidates arguably satisfy the individual criteria (Criteria 1-4). However, we could not select all of these facets, because of the set Criteria 5 and 6—the facet set has to be both "small enough" to be kept in mind and "large enough" to make an impactful difference.

Neither "small enough" nor "large enough" is precisely measurable, but we can inform our choices with field data from GenderMag [35]. Some GenderMag field reports suggest that keeping even five facets in mind at once may be more than some groups of GenderMag evaluators are comfortable with. For example, in two field studies of multiple teams of software practitioners [33, 79], most teams tended to have a favorite 2-3 facets that they used for most of their evaluation, with much less attention paid to the other facets. However, the different teams did not choose the *same* particular facets to emphasize. For example, some have focused primarily on the risk and learning style facets [33], others have emphasized the information processing facet [33], and others have emphasized the self-efficacy and motivations facets [139]. Thus, this set of facets seems to be viably small while also giving different teams the ability to focus mostly on facets that make the most sense to them. That facet set has also turned out to be large enough to be powerful [33, 139, 200], with a very low rate of false positives [35, 200]. This empirical evidence suggests that a facet set with about five elements satisfies Criteria 5-6, so we likewise aimed for about five facets for our SESMag facet set.

We ultimately chose six, namely the six facet candidates that most clearly satisfy individual Criteria 1-4 and/or contribute to group Criteria 5-6. The subsections that follow present each of these final facets, their importance to problem-solving technology's support of SES-diverse populations, and the extent to which each facet satisfies the above criteria.

### 3.1 Facet 1: Access to Reliable Technology

The first facet we selected from the candidate facets is Access to Reliable Technology. This facet includes access to technological devices and the internet, as well as reliability of that access. We selected this candidate as a facet for two reasons: (1) an individual's SES can affect the technologies they choose to use (e.g., how often they use cloud-based technology); and (2) the kind and reliability of hardware and internet an individual uses are, in turn, prerequisites to the use-cases in which an individual can participate and the quality of their user experiences in doing so.

First, consider how the technology infrastructure an individual regularly uses, such as hardware devices and internet access, varies for different SES strata [67, 78, 117, 173, 192, 203, 215]. For example, more high-SES than low-SES households in the U.S. own computers, with rates ranging from 72% (low-SES) to 99% (high-SES) [163]. The type of computer also can vary by SES. For example, some low-SES populations relying mainly on mobile phones to access the internet (e.g., [78, 163]), whereas use of tablets is much more common among high-SES households than among low-SES households [199]. In addition, whether an owned computer is shared by multiple members of the household varies by SES, with low-SES families more likely to share devices among their members, and also to use them in shared/public spaces like libraries for reasons such as internet affordability [158, 202, 215]. Connectivity with the internet varies by SES, ranging, for example, from 21% of low-SES U.S. households to 80% of high-SES U.S. households having high internet connectivity [163]. In addition to reasons of affordability, some reasons that have been reported for low connectivity include slow service and the use of older technology [67, 158]. Finally, ability to maintain one's technology varies by SES. As the technology maintenance construct argues, the high cost of maintaining digital resources disproportionately disadvantages lower-SES individuals [76]. All of these infrastructure differences can influence the user experiences available to individuals in different SES strata.

These infrastructure differences across SES strata can impact *how* individuals try to use technology, *how* they learn about it, and what their *user experiences* are. For example, Yardi et al. found that although low-SES families are more likely to share devices, low-SES teens have more independence in technology than teens in high-SES families [215]. Further, low-SES teens sometimes take on more responsibility than high-SES teens to ensure the family's successful technology usage [215]. Other types of usage can differ by SES too, particularly among low-wage earners and individuals supported by the social service system [66]. Another difference reported in the context of higher education was that most low-SES students used the internet primarily for communication through email, discussion forums, and downloading music [173]. In contrast, high-SES students used the internet for a broader range of activities, including reading news, playing video games, and seeking information on goods or services [173].

These phenomena suggest the following insights into how SES can come together with individuals' technology usage:

**Insight1**: Level of familiarity with particular technologies and features will vary by SES, with high-SES populations having the most familiarity with the most modern technologies, and low-SES populations having the most familiarity with using mobile technologies for tasks more traditionally done on laptops or desktops.

**Insight2**: Supporting use of shared infrastructure (e.g., computers, internet accounts, etc.) without loss of functionality or privacy will be useful to users of all SES levels, but especially to low-SES users.

**Insight3**: Technology and websites that provide good user experiences on not only laptop/desktop but also mobile devices will support users across a wider range of SES strata than technology and websites that support only one of these platforms.

Appendix A enumerates how this facet satisfies the inclusion criteria for Criterion 1 (was a facet candidate), Criterion 2 (at least 5 different studies/papers), and Criterion 4 (has implications for technology design). We have reason to expect it to also satisfy Criterion 3 (understandability even without social science background), because occasional bouts of unreliability and loss of internet access occur regularly in technology fields, and software practitioners are likely to have experienced these issues themselves.

### 3.2 Facet 2: Communication: Literacy, Education, Culture

Appendix A includes language usage, education, and culture as separate facet candidates. Because the literature shows numerous interconnections among these three facet candidates [53, 135, 148, 170], we combined them into one facet to capture an individual's communication fluency—specifically, fluency with the language the technology uses to communicate with users.

Research from multiple countries has shown that an individual's SES is a factor in language literacy, even in their native language [3, 53, 59, 100, 113, 138]. Research further shows that literacy level in the technology's communication language matters to users' success in navigating that technology [45, 80]. Literacy is not only grammar and vocabulary; it also includes a comfort level with different structures in written communications. For example, some low-literacy individuals understand enough vocabulary and grammar to engage with reading at the level required for basic functions and needs, but not at the level of complex sentences or paragraphs such as those in newspapers (e.g., [113]). These factors suggest that technology designers need to consider the levels of fluency that diverse members of their target population might have regarding the technology's communications.

Beyond grammar, layperson vocabulary, and structure, technology's communication-related design choices also include the idioms, specialized vocabulary, cultural references, and sentence structuring that some particular technology requires for communicating with it [80]. Sometimes the technology's communications are in English, sometimes in other languages, sometimes in specialized languages involving specialist vocabularies, and sometimes reference an assumed shared culture. As an example,

Stack Overflow's pages for answering questions about spreadsheet formulas (https://stackoverflow.com/questions/tagged/google-sheets-formula) often use "techie English"—with tech-specialist vocabulary ("url", "html", "dynamic length", "iterate", "loop through") and assume shared tech-specialist experiences that obviate the need for details ("next, you flatten it"). Further, some of these pages' vocabulary and structure require college-level reading ability[1].

Education is obviously a critical factor in an individual's opportunities for literacy-learning activities. However, systemic inequities in education consistently provide education whose quality varies with an individual's and their community's SES strata, as has been shown across all age groups in over 40 countries [10, 25, 46, 49, 72, 87, 160, 190]. Inequities in education quality for different SES levels may have become even more pronounced during COVID-related school closures. As examples, the adoption of remote and hybrid education pushed some students from low-income communities to negotiate shared electronic devices, even though multiple children needed to attend different classes at once; to use only a cellphone to do homework meant for laptops; and/or to do their schoolwork in businesses' parking lots to access the internet [131, 198]. Education *about* technology is also not equitable across SES levels, such as in providing access to technology-related courses in schools [117, 119, 203]. Due to factors like these, a multitude of significant relationships between a family's SES and their child's technology-related achievements have been shown [1].

An individual's education experience is mediated by the culture with which they identify, and members of minority ethnicities and cultures are disproportionately low-SES; for example, in the US, over twice as many African-American and Latino children than White, Asian and non-Latino children live in poverty [10]. On the other hand, education research into supporting cultural diversity with technology has revealed several ways members of minority groups gain literacy education through technology, and without loss of ties to their own cultures. A grassroots example is when mothers of children experiencing life on both sides of the U.S.-Mexico border used multiple digital tools to help educate their children in English fluency while still staying connected to Mexican culture. For example, one mother had her child watch Spanish science videos, listen to English music, and play Xbox games in English, for the mother to monitor the child's bilingual and bicultural exposure and comfort level [135]. Another study showed that Massively Multiplayer Online Games (MMOGs) can serve as an effective tool for learners of English-as-a-second-language to pick up American culture in an immersive, culturally safe setting [170]. In that study, Shahrokni et al. performed a narrative inquiry into how native and non-native English speakers interacted in the game Stronghold Kingdoms, finding that through teamwork, shared goals, and socializing, players were able to gain fluency with the online community's linguistic and cultural norms [170].

These findings about communication differences across SES levels, and their interdependencies with language literacy, culture, and education, suggest the following:

**Insight4**: Technology that has been designed with mindfulness of the literacy background and education of diverse members of their target populations will be inclusive across more SES levels than technology that has not been designed in this way.

**Insight5:** Technology that is culturally flexible (e.g., can reflect the culture with which an individual is most familiar) will be inclusive across more SES levels than technology that is not culturally flexible.

Regarding the inclusion criteria, this facet satisfies Criteria 1, 2, and 4 for the final facet set, as detailed in Appendix A. We expect the language and education aspects to be understandable and directly usable (Criterion 3), but the culture aspects' relationship to communication literacy may require more social science background than some software practitioners have. Empirical research of its usage in the field will be required to answer this question.

### 3.3 Facet 3: Technology Self-Efficacy

From the candidate facets in Appendix A, we selected Technology Self-Efficacy as one of the final facets. Self-efficacy is the belief an individual has about their ability to perform an upcoming task to achieve a goal [21]. Self-efficacy is applied to a context, and here we use *technology self-efficacy*, an individual's belief in their own abilities to interact with unfamiliar technology. Self-efficacy's effects on an individual's ultimate success with a task can be numerous, including whether they blame themselves for difficulties they encounter, and their willingness to persevere in the face of difficulty and try different approaches to the problem if their first attempt fails [21]. Such effects have been shown multiple times with technology/computer self-efficacy (e.g., [179]).

[1] As per readabilityformulas.com.

Many studies have established differences in technology/computer self-efficacy tied to people's SES strata. For example, in a study conducted across 53 Norwegian schools, Hatlevik et al. found that students' technology self-efficacy was significantly influenced by their SES and prior experiences with technology [82]. As Huang et al. summarized, the interplay between an individual's technology self-efficacy and their SES level is not surprising, given the social science underpinnings in common. As Huang et al. put it, "Four … factors which may affect <self-efficacy> include attainment, experience, social persuasion, and physiological factors (Bandura, 1995). Interestingly, these four factors may all be affected by family SES as well" [82].

As Bandura explains [21], an individual's past experiences and education factor into an individual's formation of these beliefs [21]. Consistent with this point, researchers have shown that barriers to access to reliable technology (which varies by SES, as discussed earlier in Section 3.1) contribute to lower technology self-efficacy. Access and education can come together as a factor as well; for example, schools in low-SES communities may not have reliable access to technology-based learning opportunities, bringing low-SES students fewer opportunities to experience technology [117, 119, 203]. In addition, one study found that due to SES differences in access to internet and/or computers, students from low-SES families expressed lower confidence in their ability to use Information and Communications Technologies (ICTs) [193]. These experiences contribute to low-SES populations' more negative computer beliefs compared to their peers from higher-SES populations [82, 194].

As the previous paragraph suggests, the value of one facet can exacerbate another facet's effect, even when the two facets' definitions are independent. This is particularly salient in the ways technology self-efficacy varies by SES. In the example above, although the facet of technology self-efficacy is *defined* orthogonally to the facet of technology access, differences in technology access can *affect* an individual's feelings of technology self-efficacy through the mediating factor the two have in common, namely experience. Another example is the facet of people's perception of control over their environments (discussed later in Section 3.5), with the factor in common being prior successes. As Bandura's writings point out (e.g. [21]), an individual's history of successes helps to determine their self-efficacy with similar future endeavors. Likewise, an individual's past successes at controlling technology outcomes affect their perceived control over future technology outcomes, as we discuss in Section 3.5. These facets individually or in combination may point some low-SES individuals toward a sense of inevitable failure in their computing tasks.

These findings suggest (at least) the following insights for a more socioeconomic-aware HCI:

> **Insight6:** Because technology self-efficacy varies by SES, people in different SES strata will have different expectations as to whether failure is inevitable with complex technology tasks.
>
> **Insight7:** Upon encountering difficulties or other negative consequences with technology, individuals' tendency to blame themselves vs. to blame the technology will vary by SES.
>
> **Insight8:** Because technology self-efficacy varies by SES, problem-solving technology that clearly states where/how to get help if needed will support individuals across a wider range of SES strata than technology that does not provide this information.

Regarding the facet inclusion criteria from the beginning of Section 3, Appendix A enumerates how self-efficacy satisfies Criteria 1, 2, and 4. Regarding Criterion 3 (understandability), self-efficacy's usage as a GenderMag facet[1] provides evidence of its understandability in literature documenting numerous software practitioners successfully using GenderMag's self-efficacy facet (e.g., [33, 139, 200]). Technology self-efficacy's usage with GenderMag also provides encouraging evidence that this facet may contribute strongly to the set Criterion 6 (impact). For example, in one study of Open-Source Software's (OSS) inclusivity issues, the self-efficacy facet helped to identify 88% of the inclusivity issues that the OSS practitioners found [139]. In another study of fixing inclusivity errors in a search engine's interface, self-efficacy inspired the design changes behind half of the fixes [200].

### 3.4 Facets 4 and 5: Attitudes toward Technology Risks, and toward Privacy and Security

Several of the candidate facets in Appendix A pertained to elements of risk that can arise in using technology. Two of these technology risks—privacy and security—are so important, an entire subfield (or two subfields by some counts) investigate(s) them. However, privacy and security are not the only types of technology risk that are particularly salient to low-SES individuals. Thus, we collected elements pertaining to risk from all the candidate facets in Appendix A, including privacy and security.

[1] Technology self-efficacy is one of the five GenderMag facets because individuals with different genders tend to cluster around different technology self-efficacy values [179].

We started out combining these risks into one facet that we named "Risk, Privacy, and Security" to keep explicit that Privacy and Security are not the only types of risks that should be considered in bringing socioeconomic awareness into HCI. However, a field test [2] showed that the facet was "too big,"—i.e., evaluators so focused on Privacy and Security they did not consider other types of risk. As a result, we broke it into two facets: "Technology Privacy and Security" and "Attitude toward Technology Risks"

Starting with the Technology Privacy and Security facet, some fears common across all socioeconomic strata relating to privacy/security are the risk of identity theft, of online financial fraud, and of hackers who might steal or take over information such as passwords [78, 180]. For example, in a recent Pew Research study [18], 28% of Americans say they have suffered from someone putting fraudulent charges on their credit or debit card, someone taking over their social media or email accounts, or someone trying to open a credit line or get a loan using their name.

However, some demographic groups that are disproportionately in low-SES strata [10] worry even more about such risks, and for good reason—such risks materialize for them more often. For example, Black adults have had someone take over social media or email accounts at almost three times the rate of White adults and are also more likely than White adults to have had someone attempt to apply for credit or a loan using their name [18]. In a study of undocumented immigrants, many participants were hesitant to post images of themselves on Facebook, for fear that they would be recognized and then arrested or deported [78].

Another reason for heightened caution about privacy and security has been the higher prevalence of police and government surveillance in low-SES communities. In some low-SES communities, surveillance is constant and brings disproportionately heavy consequences for people in those communities (e.g., greater rates of arrest and police- or government-related deaths for Black and Hispanic people, who are disproportionately low-SES [10]). Low-SES individuals can also feel particularly susceptible in other ways that arise less often than with higher-SES individuals' privacy and security concerns. One group of such issues arises because of the prevalence of shared devices among low-SES individuals (e.g., [215]), as mentioned in Section 3.1. In these situations, individuals may not trust the people who share their devices to not view their browsing histories, data, apps, etc. [165]. These situations, which vary with SES, are reflected in attitudes toward privacy and security that likewise vary with SES. For example, Black Americans (disproportionately lower-SES) are more likely than White Americans to worry that the government is tracking them online, and Black and Hispanic Americans (also disproportionately lower-SES) are more likely than White Americans to be concerned about what law enforcement officials, employers, and family and friends know about them [18].

The steps individuals take to guard against these risks can themselves further impact their user experiences. In addition to common security/privacy practices like phone locks and passwords, some low-SES individuals avoid technology risks by avoiding engaging with technology whenever possible [29, 165]. Another tactic reported in a study in three South Asian countries was to delete regularly and extensively. In that study, some participants safeguarded privacy by removing entire threads or histories and others deleted specific chats, media, or queries [165]. As one participant in that study put it, "Privacy is not for me, it's for those rich women" [165]. Another study of 28 low-SES New York youth likewise found that its participants were very cautious with their online information [120]. Among their privacy measures were asking friends to remove social media tags identifying them, because they feared consequences such as family drama or compromised employment [120].

Moving beyond privacy and security to the Attitude toward Technology Risks facet, other technology-related risks also vary by SES. For example, a risk we have already discussed (Section 3.1) is that of unreliability—when using less reliable devices or less stable internet connectivity, the risk is that technology will fail before an individual can complete the task they are trying to accomplish with it [196, 215]. Even when the technology holds up, many low-SES individuals risk experiencing disappointing outcomes from using technology. For example, online job searches tend to support professional and highly educated job seekers, but not jobs more commonly held by low-SES individuals [58]. Other outcomes have been reported to be especially prevalent since the 2016 U.S. election, due to an increasingly xenophobic online atmosphere fueled by political tensions. As such, many low-SES participants reported not engaging in public online discussions, for fear of doxing or online harassment [78].

Finally, alone or in combination with many of the above risks is the risk of wasting time without gaining benefit. Low-SES families sometimes must make ends meet by working multiple jobs [8, 14, 44]. They also tend to have multiple, competing demands on their time, such as requiring extra time to acquire basic necessities for their children. For example, their time often goes to complicated logistics due to use of public transportation and time-consuming dealings with bureaucracies for income, food stamps, etc. [216]. Thus, when they perceive risk of failure due to one of the risks above, they may also perceive a risk of wasting their limited time on work that will not succeed.

These findings on privacy/security and other kinds of technological risks imply the following:

> **Insight9a, b:** Individuals' perceptions of technology features as being risky to use, due to SES-varying perceptions of (Insight9a:) privacy risks, security risks, (Insight9b): risks of lost time, risks of unfavorable outcomes, etc., will vary with their socioeconomic status.
>
> **Insight10a, b:** Individuals' actual usage of features in the technology will vary with their socioeconomic status in the same way as their perceptions of its (Insight10a and Insight10b): risks.

Appendix A enumerates how the Privacy/Security facet and Risk facet satisfy Criteria 1, 2, and 4 in the final facet set's inclusion criteria. As the Appendix shows, even without considering the privacy and security aspects, the risk aspect satisfies the final facet set inclusion criteria. As with the self-efficacy facet, insights into this facet's ability to fulfill Criteria 3 and 6 can be gleaned in risk's presence as a GenderMag facet [35]. In the GenderMag context, the Risk facet (which includes Privacy/Security in the GenderMag version) has been impactful (Criterion 6), which suggests that it has been understandable enough for software practitioners to use (Criterion 3). For example, in one study of Open-Source Software's inclusivity errors, software practitioners used the risk facet to help identify 71% of the inclusivity errors they found [139]. In another study of fixing inclusivity errors in a search engine's interface, software practitioners used the risk facet as a foundation behind *all* of the inclusivity error fixes they devised [200].

### 3.5 Facet 6: Perceived Control and Attitude Toward Authority

The sixth facet is a combination of two of the facet candidates, Perceived Control and Attitude toward Authority. Although each of these facet candidates varies with an individual's SES and each has potential impacts on how an individual interacts with technology, we chose to combine them into one facet in that each of them interrelates with the other in ways that are difficult to separate.

Perceived Control refers to an individual's belief that they can influence future events [140]. The degree of agency and control an individual feels over their life tends to vary with SES; for example, lack of agency and control over one's life are frequently reported feelings by low-SES people [147, 155]. An individual's perception of control over their lives also interacts with their attitudes toward and behaviors with authority figures, and these too vary by SES. For example, few low-SES individuals are in positions of power, which relates not only with their perceptions of control, but also comes with a requirement to comply with the dictates of authority figures [69, 81, 188]. Consistent with this reality, decades of research have strongly correlated low-SES individuals' perceived lack of control with a tendency to be accommodating to authority figures (e.g., [186, 188, 213]). Further, due to various societal hierarchies, low-SES individuals do not interact with authority figures as their equals and are also less likely than higher-SES individuals to be overtly critical of authority figures [184].

An individual's perception of control over their outcomes and attitude toward authority directly ties to their behaviors with and around technologies. For example, Petro et al. reported SES differences in students' coping mechanisms when technology disruptions arose [147]. They point to a long-standing result from sociology that lower-SES students are more likely than higher-SES students to require self-reliance of themselves rather than asking for help in the classroom because of negative past experiences of being judged by teachers or peers. Their results showed that this SES difference also applied to technology disruptions with class assignments, and that SES was associated with significantly greater negative consequences as a result and less efficacious beliefs about handling future similar situations [147]. Other studies have reported similar phenomena [39, 87]. Further, some people regard computers as "authorities" themselves or are willing to grant computers authority over them in some circumstances [23, 88, 106, 172]. This attitude, when coupled with an attitude of deference to authority, suggests that low-SES computer users might be more accepting of negative computer outcomes (e.g., web pages with greyed out "submit" buttons, paths leading to error messages, etc.) than people who are more critical/questioning of authority.

A potentially contributing factor to users' acceptance of negative computer outcomes is a form of "victim-blaming" known as the fundamental attribution error—a tendency to overestimate an individual's responsibility for a negative outcome that overlooks the impact of situations [176]. This tendency is related to the Just World Hypothesis (JWH), a type of system-justifying belief in which good things happen to good people and bad things happen to bad people [9]. Due to systemic disadvantages and accompanying feelings of lack of control and powerlessness, low-SES individuals are more likely to hold system-justifying beliefs that validate, internalize, and legitimize existing hierarchies, including the JWH [86]. Interestingly, system-justifying beliefs like the JWH are positively correlated with an individual's perception that they have control over their future and that advancement is possible, which brings mental and physical health benefits such as increased self-esteem [122, 207]. Thus, low-SES individuals can receive psychological benefits from perceiving society's distribution of outcomes as fair—even if that distribution puts themselves at the bottom [122]. Although multiple theories consider why some low-SES people hold system-justifying beliefs, most theories agree that system-justifying beliefs serve as a response to an unfair social system, i.e., low perceived control [69, 90, 213].

The combination of perceived control over outcomes in life, attitudes to authority, and system-justifying beliefs can interact with technology usage in nuanced ways. For example, these attitudes can affect people's "planfulness"—willingness to take actions now that are expected to bring benefit in a more distant future differ [24, 54, 125]. For example, Laurin et al. found that low-SES individuals were more likely to work toward future goals when they believed that rewards would be distributed fairly, i.e., they would get what they deserve. On the other hand, higher-SES individuals worked less while having the same belief that awards would be distributed fairly [104] (e.g., Figure 2). These perceptions can interact with the problem-solving technology that is our scope, because much of problem-solving involves planning ahead and investing time in work now that is expected to pay off later [26].

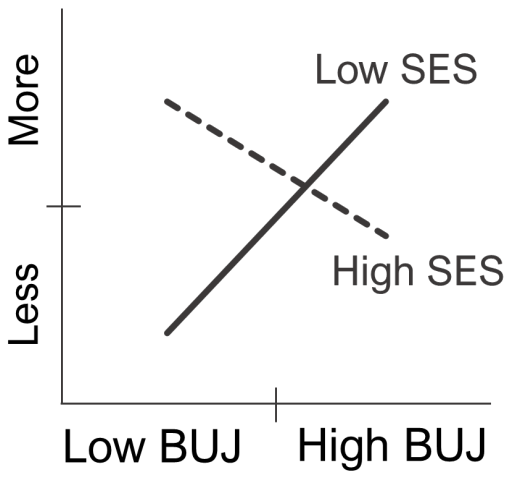

Figure 2: Laurin et al.'s participants' commitment to do well on their second test (y-axis) after having done poorly on the first test, according to their fairness beliefs (x-axis) [104]. (BUJ = Belief in Ultimate Justice, in Laurin et al.'s study.)

From these foundations on perceived control and attitude towards authority, we derive the following:

**Insight11:** Individuals' acceptance of technology's errors, dead-ends, or unfavorable results as outcomes to which they have no recourse will vary with their socioeconomic status.

**Insight12:** Individuals' willingness to engage in time-intensive future-oriented activities to overcome a problem involving technology—such as by investing time to learn complex feature sets—will vary with their socioeconomic status.

Appendix A enumerates how this facet satisfies Criteria 1, 2, and 4 in the final facet set's inclusion criteria. We also expect the facet to satisfy Criterion 3, because many people have experienced the frustrations of lack of control over some aspects of technology, but field investigation is still needed to gather empirical evidence of this criterion being satisfied.

Table 1 summarizes the six facets we have just discussed.

Table 1: Summary of the six facets.

| Facet | Potential Impacts on Technology Usage |
|---|---|
| **Access to Reliable Technology:** Amount of access to reliable devices with reliable internet access. Includes owning vs. sharing vs. depending on public devices [27, 47, 48, 57, 67, 68, 78, 101, 105, 117, 118, 119, 142, 158, 173, 192, 193, 202, 203, 209, 215].<br>Facet's value range: Low to high access/reliability. | Example potential impact:<br>Consistency of access to reliable technology that is well-suited to the task at hand affects how often, when, and how an individual uses technology.<br>Low-SES examples:<br>Only 62% of households of people living in poverty have computers, and of those, fewer than half have internet access [67].<br>Some low-SES individuals depend primarily on their mobile devices to access the internet [78]. |
| **Communication: Literacy/Education/Culture:** Fluency of communicating with system's target language. Includes vocabulary and knowledge of its idioms, use of cultural references, jargon, and quality/level of education in the target language [5, 25, 30, 53, 56, 61, 73, 80, 107, 112, 113, 127, 133, 135, 142, 148, 150, 170, 195].<br>Facet's value range: Low to high fluency with system's communications, due to literacy, education, and/or culture. | Example potential impact:<br>Users whose backgrounds did not provide the particular cultural references, literacy level, and language usage upon which the technology relies may be less successful in using the system than users with backgrounds that did.<br>Low-SES example:<br>Schools attended by mostly low-SES individuals tend to offer lower technology-related education than those attended by mostly higher-SES individuals [87]. |
| **Technology Self-Efficacy:** User's perception of their own ability to perform new/unfamiliar technology and computer-based tasks [1, 167, 187, 193, 194, 215]<br>Facet's value range: Low to high technology self-efficacy. | Example potential impact:<br>Individuals with lower technology self-efficacy who run into technology barriers may not persevere in trying to overcome them [21] because they may underestimate their ability to succeed.<br>Low-SES example:<br>Some low-SES individuals have fewer opportunities to use technology, resulting in lower technology self-efficacy and the effects that accompany it [21, 193]. |
| **Technology Privacy and Security:** Tolerance of potential risk of privacy loss, device security, etc. [18, 29, 78, 120, 124, 165, 180, 189, 196].<br>Facet's value range: Low to high tolerance for privacy/security risks. | Example potential impact:<br>Users averse to taking privacy risks (revealing identity or location) will be unlikely to use some kinds of technology.<br>Low-SES examples:<br>SES is a predictor of having had contact with the criminal justice system, and individuals who have had contact with the criminal justice system avoid surveilling institutions, including medical, financial, labor market, and educational institutions [29, 120].<br>Low-SES individuals are more likely to mistrust online banking due to institutions' potentially questionable usage of their financial information or risk of personal identity theft [196]. |
| **Attitudes toward Technology Risks:** Willingness to take technological risks (beyond privacy/security) ranging from using unfamiliar software features, risk of wasting time on fruitless efforts, risk of financial problems, etc. [10, 77, 78, 120, 124, 196].<br>Facet's value range: Low to high tolerance for technological risks. | Example potential impact:<br>Users averse to taking risks will be unlikely to use some kinds of technology.<br>Low-SES examples:<br>Low-SES individuals are less likely to take financial risks than higher-SES individuals [77, 196].<br>Individuals in racial/ethnic minority groups, who are statistically likely to be low-SES, have concerns about online harrassment and online bullying [78]. |
| **Perceived Control and Attitude toward Authority:** Users' belief in their own ability to exert influence over technology's positive or negative outcomes and interactions. Can relate to their attitude toward authority figures and the technology's outputs as an authority [38, 39, 52, 69, 86, 87, 91, 97, 103, 104, 122, 147, 151, 155, 156, 184, 186, 188, 207].<br>Facet's value range: Low to high perception of control, . | Example potential impact:<br>Some public discourse maintains that people can avoid police involvement by behaving lawfully; however, members of various marginalized groups are often disproportionately targeted by violence, which challenges the notion of having control over such outcomes in physical and online worlds [120].<br>Low-SES examples:<br>To exert the control they can, many low-SES individuals are extremely careful about providing any personal details whatsoever to any online system [120].<br>Some low-SES students are more likely to solve technological problems independently because they do not expect a helping relationship with authority figures, whereas some high-SES students are more likely to |

## 4 FROM FACETS TO ACTIONABILITY

As we have emphasized, our goal is bringing *actionable* socioeconomic awareness to HCI practices. Fortunately, the extensive foundations by the researchers whose work we synthesized in Sections 2–3, enable us to achieve such actionability in (at least) the following five ways:

*A socioeconomic-aware vocabulary*: HCI practitioners can use the six SES facets as a vocabulary in design discussions, and in so doing, can gain expressive power. Section 4.1 details this possibility.

*A set of socioeconomic-aware heuristics:* Besides producing facets, Section 3 also produced 12 insights. Because these insights are evidence-based, we believe that HCI practitioners can use them directly as heuristics for design. However, this premise has not been field-tested, so its utility as a set of design heuristics is an open question.

*Socioeconomic-faceted multi-personas:* One way to bring the facets to life is by using them as the basis of (multi-)personas, which we illustrate in Section 4.2.

*SocioeconomicMag:* Under the InclusiveMag meta-method, a natural by-product of the facets can be a systematic inclusive design process for socioeconomic awareness. Following this meta-method, we have created SocioeconomicMag (Socioeconomic Inclusiveness Magnifier), as explained in Section 4.3.

*A socioeconomic-aware design "starter set":* Identifying SES-inclusivity problems using one of the mechanisms above leads to a need to *fix* those problems. Toward this end, we harvested a set of design recommendations from the literature, which we present in Section 5.

### 4.1: The facet endpoints as an expressively powerful vocabulary

As we have explained, the facets are types, and like most types, each facet type represents a potentially infinite range of possible facet values. Note also that the SES facets are ordinal (linear); Table 1 shows each facet's two endpoints. For example, the two endpoints of the Access to Reliable Technology facet are "low" and "high" access/reliability. By construction (recall Criterion 2) for each facet, one endpoint is disproportionately common among lower-SES individuals, and the other disproportionately common among higher-SES individuals (Figure 3(top)).

Treating these values as endpoints provides an order of magnitude more *expressive power* to the facets than values alone can do. The source of this power is that technology features that *simultaneously* support *both endpoints* of a facet range will also support all the other facet values within that range. For example, if a technology product is *simultaneously* resilient to some users' unreliable/rare wifi access and others' highly reliable, fast wifi, it also supports other users' somewhat spotty wifi, slow wifi, occasional wifi outages, and so on. Thus, because of the requirement for simultaneous support of both endpoints, the HCI practitioner's task is reduced from considering each one of the 6 facet types' $n$ values—$6n$ values—to 12 facet values.

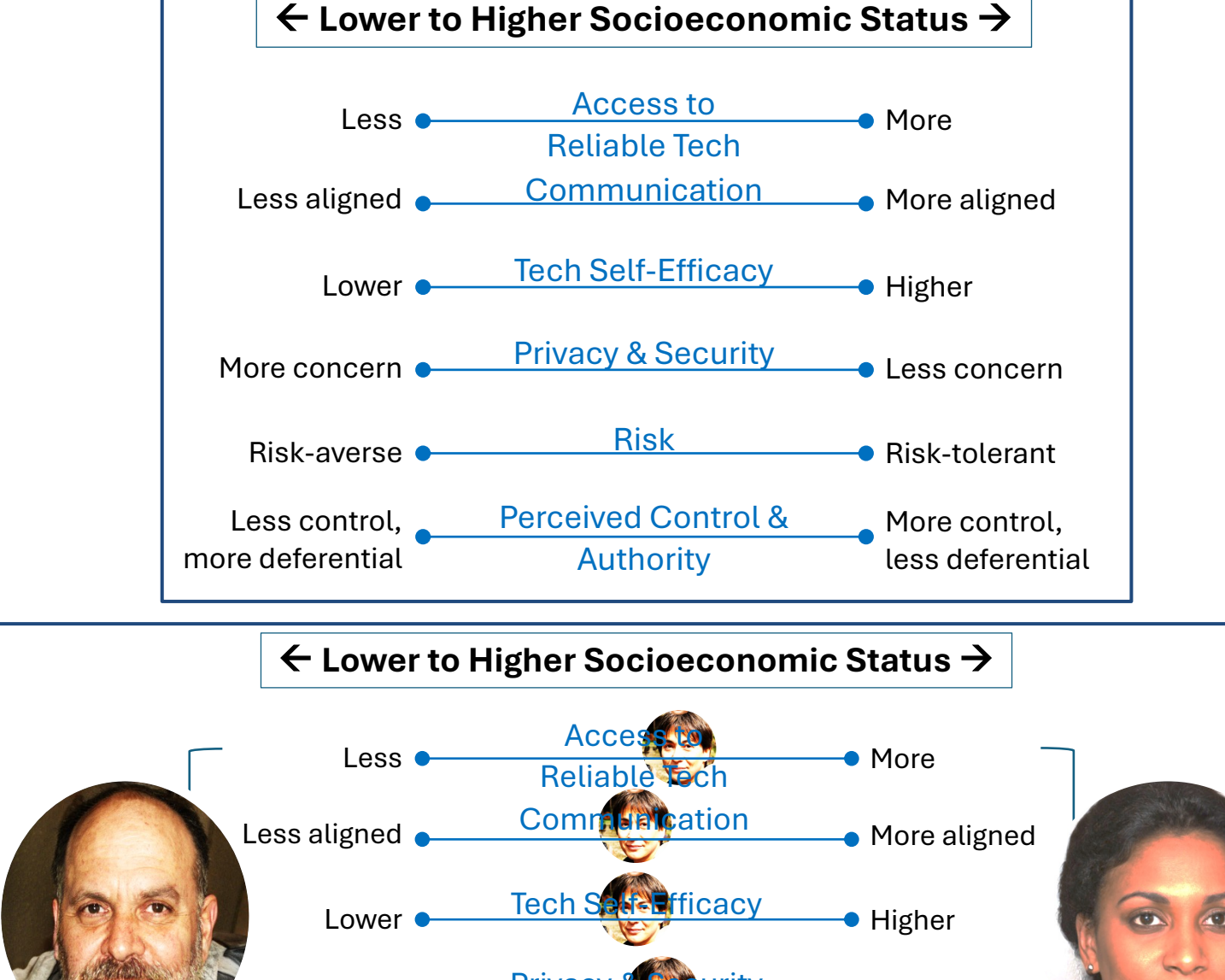

Figure 3: (Top): Each facet's range of possible values. Those on the left are more common among lower-SES than higher-SES individuals, and vice versa for those on the right. (Bottom): Dav, the lower-SES multi-persona, has the facet values at the leftmost endpoints, and Fee, the higher-SES multi-persona, has the facet values at the rightmost endpoints. Ash (small pictures at various positions) has the mix of facet values shown.

One actionable way to leverage this power is to simply use the facet endpoints as a vocabulary. For example, instead of raising identity-based questions like "will low-SES people find this error message actionable?", practitioners can leverage the facet endpoints as a vocabulary—e.g., "will people who are highly deferential to authority find this error message actionable?" In fact, Microsoft reported using the analogous GenderMag facets [35] as a design vocabulary in just this way for some of their products [36]. Besides being a lightweight use of these facets, the facets-as-vocabulary approach can mitigate potential political problems and/or seeming to stereotype users by their SES identities. For example, in the Microsoft example, the use of facets-as-vocabulary enabled considering gender-inequitable features "<without> talking about gender" [36].

### 4.2: Embedding the facets in multi-personas

For HCI practitioners' whose work practices include personas, we have created two multi-personas to house the facets' endpoints. We also created a third multi-persona to make explicit that many individuals do not perfectly align with one set of endpoints or another, but rather have a mix of facet values. The three multi-personas are Dav, who represents facet endpoints disproportionately common among lower-SES individuals; Fee for endpoints common among higher-SES individuals; and Ash, who has a mix of values (Figure 3 (bottom)).

*Multi-personas* are persona templates, originally created for GenderMag personas [85], in which certain portions are fixed and in other portions practitioners fill-in-the-blanks as needed. Thus, a multi-persona can generate infinite variations of the same template, that vary only in controlled ways. In the SESMag multi-personas, the facet values are fixed, but customization is enabled for most non-facet related information, such as the persona's photo, age, ethnicity, job/trade/profession, education, or anything else that places this persona in a position to use a particular technology product. For example, Agarwal et al. did a field trial of the Dav multi-persona. In that study, HCI practitioners were working on a learning platform for community college students, so they customized Dav to be a community college student who also worked fulltime in a grocery store [2]. Figure 4 shows Dav; all three personas are given in Appendix B.

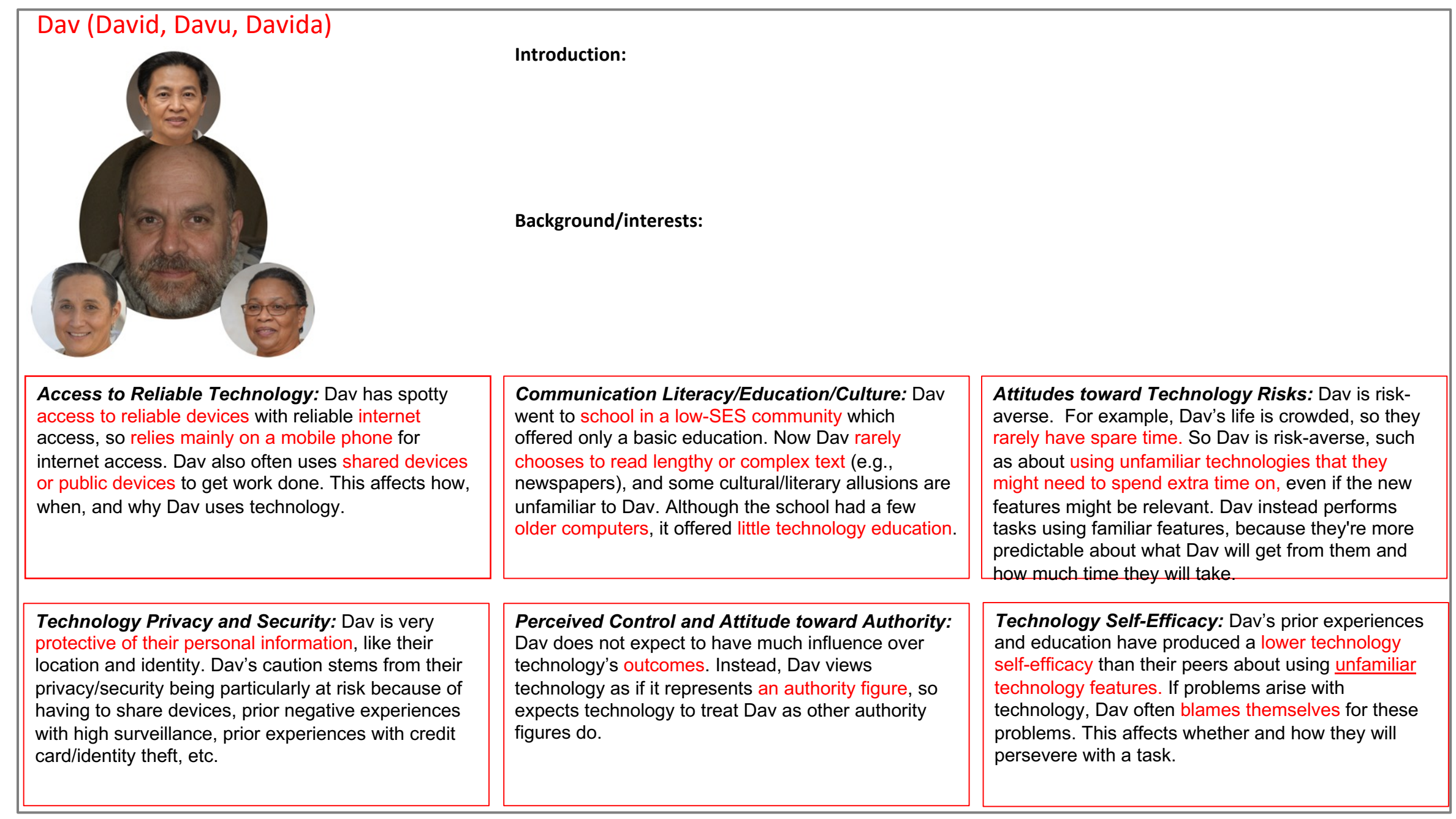

Figure 4: "Dav", the customizable multi-persona whose facet values are at the lower-SES endpoints. Appendix B shows the customizable multi-personas for "Fee", whose values are at the higher-SES endpoints, and "Ash", who has a mix of facet values.

### 4.3: Leveraging the facets and personas in a systematic method → SocioeconomicMag

Readers familiar with the InclusiveMag meta-method may notice that our process in this paper aligns closely with InclusiveMag's process. InclusiveMag [126] is a method for creating systematic inclusive design methods. It was used to create the GenderMag (Gender-Inclusiveness Magnifier) method, which has been extensively validated and field-tested [33, 35, 36, 79, 84, 85, 139, 200]; the AgeMag method [123]; and a variety of small academic inclusive design methods [34, 126].

As Figure 5's left dotted box enumerates, we have already completed most of the steps for inclusivity researchers to create *SocioeconomicMag*, a new systematic inclusive design method for analyzing a product's socioeconomic inclusiveness. In Section 1, we explained the diversity dimension of interest (SES). Sections 2–3 described the research resulting in the SES facets and facet values, and in Section 4.2 we created personas to bring the facets to life.

We now select a technology type of interest (top left in the figure), and an analytic process to specialize using the facets and/or personas (bottom right of the first dotted box). These two choices are related, because the choice of analytic process needs to work well for the chosen technology type. For the analytic process, we choose the Cognitive Walkthrough (CW) [206], both because of its high accuracy (very low false-positive rate) in the field [114] and because of its strong empirical results with other InclusiveMag-generated methods [35, 79, 139, 200]. Since the research behind the CW comes from psychological research on problem-solving [206], we choose problem-solving technologies (e.g., spreadsheets, debugging tools, decision-support tools, information-seeking aids, etc.) as our technology type to set the scope of our method.

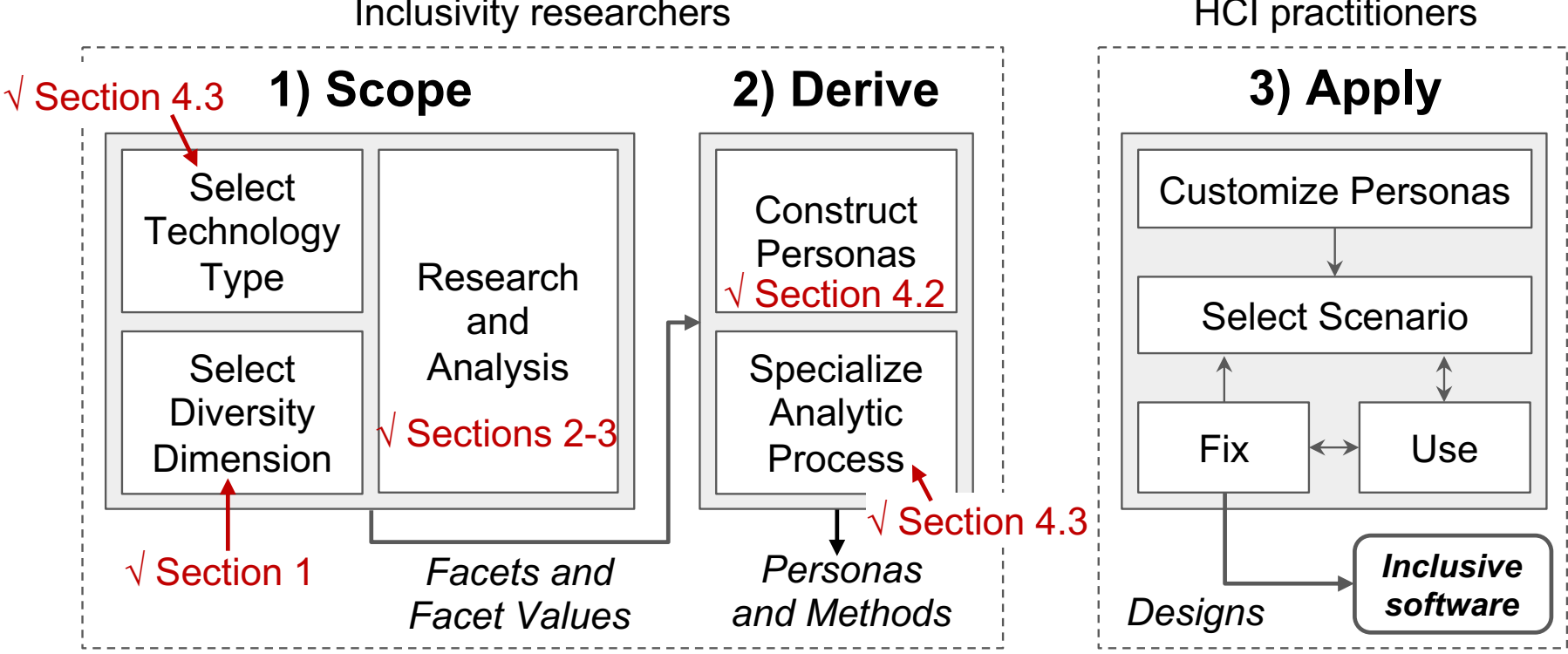

Figure 5: The InclusiveMag meta-method's [126] three parts. Inclusivity researchers perform first two, and HCI practitioners perform the last one. Magenta notes show where this paper addresses each of the inclusivity researchers' steps.

With a CW, practitioners "walk through" each step of a user task, answering questions at each step about how a user might experience the software at that moment [206]. For SocioeconomicMag, we specialize the standard CW by replacing generic versions of "the user" with whichever of our personas a practitioner instantiates, and explicitly call attention to the six SES facets, as follows:

---

**For each subgoal:**

Will <persona> have formed this sub-goal as a step to their overall goal (yes/maybe/no)? Why?
Which, if any, of <persona's> facets did you use to answer the question? <facet list> Why?

**For each action:**

1. [BEFORE ACTION] Will <persona> do this (yes/maybe/no)? Why?
   Which, if any, of <persona's> facets did you use to answer the question? <facet list> Why?
   What in the UI helped/confused <persona> in this step?
2. [AFTER ACTION] If <persona> does this, will they know they did the right thing and are making progress toward their goal (yes/maybe/no)? Why?
   Which, if any, of <persona's> facets did you use to answer the question? <facet list> Why?
   What in the UI helped/confused <persona> in this step?

---

Each "maybe" or "no" answer to these questions is a potential usability bug for that persona. Further, if the reason is that the technology failed to support one of the persona's facet values, then that is an inclusivity bug—because the association of that usability bug with a facet value means the bug will *disproportionately* affect individuals with that facet value.

These steps complete the process of creating the new SocioeconomicMag inclusive design method, leaving the application of it to interested HCI practitioners (right section of Figure 5).

## 5 HOW THE LITERATURE'S DESIGN RECOMMENDATIONS MEET THE FACETS

We have shown five ways HCI researchers and practitioners can use the facets to take action, and many of these actions result in identifying SES inclusivity problems in technology. But once they identify an SES inclusivity problem, what should they do to *fix* it?

The literature offers the beginnings of answers to this question, in the form of recommendations on how to design technology for some low-SES subpopulations in specific situations. Below we present the design recommendations we located, ordered by the six facets, and Figure 6 summarizes.

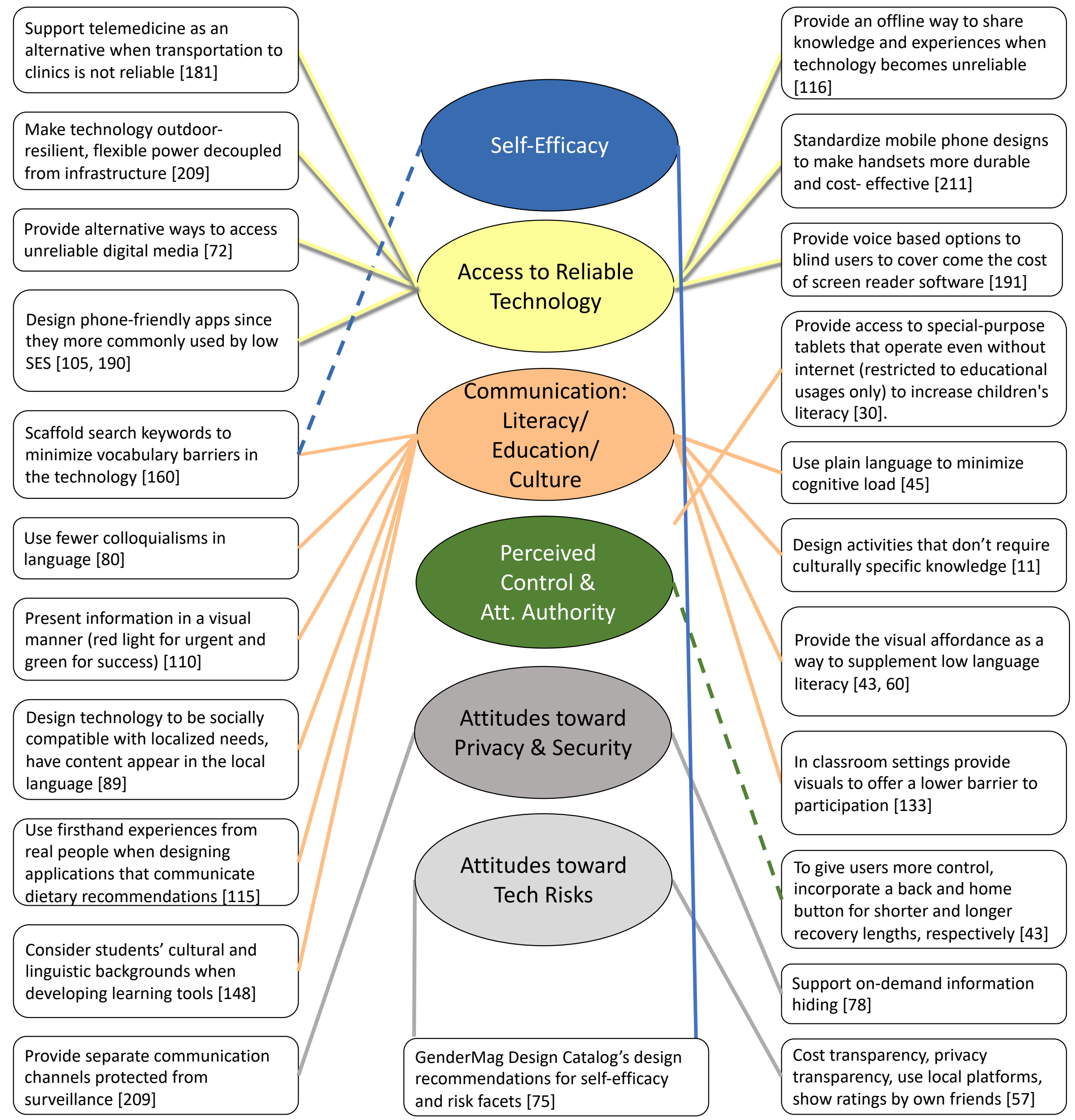

Figure 6: Solid lines denote a direct connection from the referenced paper to the facet, and dashed lines denote a connection from the referenced paper to our facet-derived insights from Section 3.

### 5.1 Design Recommendations for Facet 1: Access to Reliable Technology

We located seven design recommendations for SES-oriented situations pertinent to the Access to Reliable Technology facet. The first is a collection of recommendations specific to supporting individuals experiencing homelessness, for whom access to reliable technology is often very problematic [209]. This collection recommends that technology should be designed to have low cost; should be outdoor-resilient (e.g., against wet and cold); should have access to flexible non-infrastructural energy sources (e.g., solar); and should have 24/7 reliability for emergency use [209]. In addition, Le Dantec and Edwards' research revealed that cell phones were the preferred device for many individuals experiencing homelessness—both as a communication device and a symbol of belonging to society [105]. Therefore, they recommended that software applications be designed to work on cell phones. More generally, Wyche and Murphy's recommendation expanded on reducing costs, and recommended standardizing mobile phone designs. They argue that this would result in more durable and cost-effective phones, which can in turn support access by individuals across a wider range of SES strata [211].

Some design recommendations, rather than focusing on increasing access to or reliability of a technology, focus on providing *alternatives* to a technology that is failing an individual. For example, Vashistha et al.'s research into low-SES blind people recommended designing voice-based forms of social media. This recommendation not only relates directly to the Access to Reliable Technology facet by providing an alternative to expensive screen reader software, it also added to participants' social skills [190, 191]. Marcelino-Jesus et al. recommended providing both online and offline spaces for students to access and share knowledge, and Hebert et al. recommended making physical (offline) copies of information available [72], to support individuals' variable access to reliable online technology [116]. Another investigation considered using cell phone technology to mitigate failures in transportation technology. Specifically, in an investigation into how low-income communities access and receive healthcare, Tang observed that low-income communities are more likely to have access to communication technologies like phones, than to transportation technologies like cars or reliable public transportation. Based on this observation, Tang recommended that telemedicine be made available as an alternative to in-person doctor appointments, to support diverse individuals' differing levels of access to transportation technologies that are reliable or that can fit their scheduling constraints [181].

These recommendations suggest that designs for varying SES levels should not assume that individuals will have access to reliable device or infrastructure. Based on these design recommendations, we propose the following heuristics:

Heuristic1: Design for variable and shared device access.

Heuristic2: Support shared infrastructure without sacrificing privacy.

### 5.2 Design Recommendations for Facet 2: Communication: Literacy/Education/Culture

We located eleven recommendations for the Communication: Literacy/Education/Culture facet. The eleven recommendations fell into three categories: recommendations about English language/vocabulary inclusivity, recommendations on visuals and imagery for inclusivity, and recommendations about cultural inclusivity.

Some technologies use advanced English vocabulary or structures to communicate, so some recommendations relate to reducing literacy barriers by maintaining simplicity in the language used in the technology (English in these examples). For example, Guo's study into how non-native English speakers learn computer programming revealed barriers in reading and writing code, as well as understanding technical vocabulary [80]. Note that two languages erected barriers in Guo's study: English language, and technical vocabulary. A third literacy barrier also arose in Guo's study: English cultural references. Guo recommended using simplified English, including simple sentence structure, fewer colloquialisms, and fewer references to English culture-specific topics [80]. Cheung likewise recommended using plain language in professional and technical software interfaces, pointing out that plain language can help minimize the cognitive load—especially among non-native English speakers—to use these interfaces [45]. Instead of reducing such literacy barriers, Kaur et al. [89] and Maitland [115] recommended outright removing some of them by having technology communicate in the user's local language.

Rather than reducing or removing literacy barriers, some researchers recommended technology-based scaffolding to help users climb over them. One example was Breazeal et al.'s education-oriented tablets, which they distributed to children from under-resourced schools. These tablets had built-in capabilities to overcome several technology access issues (i.e., they could work without an Internet connection, supported remote administration, and limited usage to educational content), so that the tablet could reliably be used by the children for their education activities. Breazeal et al. found that children who used the education tablets showed stronger literacy skills at the end of the study than similar-grade peers from the same school who were not given the tablets. Thus, they recommended providing similar low-cost, educational mobile devices to low-SES children for improved literacy outcomes [30]. Another example related to vocabulary scaffolding. Roshan et al. (also discussed in Section 5.3) reported some low-SES parents running into literacy barriers in finding useful online educational resources because of vocabulary mismatches in internet searches. Therefore, the researchers recommended investigating the search keywords parents use in education-oriented searches, and creating custom tag labeling structures for online educational datasets that match parents' keywords [160].

A third way to reduce literacy barriers is to replace use of written language with use of appropriate visuals and images. For example, Chaudry et al.'s study on designing a mobile health application for low-literacy and low-technical-literacy users, found that their participants favored visual elements such as non-text-based graphical user interface widgets and large radio buttons over text [43]. Similarly, Liu et al.'s qualitative study involving a prototype mobile application to support communication between doctors and low-SES parents caring for high-risk infants, recommended incorporating recognizable visualizations [110]. By communicating a red light for high alert and green light for success, researchers were able to mitigate literacy and language barriers that the parents faced upon communication with their pediatric providers. Nacu et al. also recommended using more visual interactions. When teaching low-SES Latino students on an online platform, they decided to use visual reactions, such as emojis, instead of open-ended textual commenting. They reported that this decision lowered barriers to participation and improved the fit to the Latino students' cultural norms [133].

The literature offered several recommendations centered on cultures. For example, in a study of information and communication design in rural, socioeconomically underdeveloped Indian villages [60], Dutta and Das recommended using community-generated visual elements that are culturally appropriate. In a study of African-American students who lived in low-SES and mid-SES neighborhoods, Rader et al. also recommended that technology take into account individuals' cultural and linguistic backgrounds [148]. Their tutoring tool was cognizant of African American communities' literacy conventions in both African-American English and in narrative storytelling. As a result, the tool helped low-SES and mid-SES African-American children pick up "school English" literacy by enabling the children to connect to the concepts in their own dialect and culture. In an investigation of health-related behavioral change technologies, Maitland observed low-SES families' motivations to make positive dietary changes [115]. In order for health messages to support *actionable* change instead of being designed to simply motivate change, Maitland recommended that health messages be delivered in a community-based manner [115]. For example, when trying to convince parents to introduce healthy foods to a child, Maitland found that a forum with firsthand experiences is preferable to a generic website that lists potential strategies. Finally, Ames and Burrell suggested avoiding cultural mismatches by designing technology with less culture-specific knowledge, such as having a treasure hunt instead of a Harry Potter activity in Minecraft [11], thereby removing barriers to participation based on an individual's prior readings or social activities.

These recommendations indicate that inclusive communication requires more than presenting information in a conventional textual format. Thus, to accommodate the diverse literacy, linguistic and cultural needs of users across SES levels, we propose the following heuristics:

Heuristic3: Use plain language, minimize jargon and culture-specific references and avoid literacy barriers.

Heuristic4: Provide alternatives to text-based communication.

### 5.3 Design Recommendations for Facet 3: Technology Self-Efficacy

We located two design recommendations relevant to supporting SES-diverse users' Technology Self-Efficacy facet values. One relates to Insight6. Recall that Insight6 relates SES to technology self-efficacy, and points out that users with low technology self-efficacy who have experienced failure with technology before may expect that failing is inevitable when attempting technology tasks that seem complex.

This expectation can act as a self-fulfilling prophecy which then reinforces the expectation, forming a vicious cycle. Breaking the cycle is thus a key strategy for addressing Insight6-type issues. For example, in an investigation into how parents in economically depressed communities find learning resources for their children, Roshan et al. recommended that learning resource search technologies should include search scaffolding, because low-SES parents were not using the same search vocabulary as the technology expected [160]. We hypothesize that support for such scaffolding will help support low-SES individuals' technology self-efficacy, because the design recommendation may prevent users from failing to locate the learning resources they seek.

The other design recommendation is a collection of recommendations about self-efficacy-related design remedies that were revealed for GenderMag's self-efficacy facet. The GenderMag project has an emerging design catalog that enumerates real-world design remedies by facet, including real-world usage and/or evidence behind each design remedy [75]. These remedies range from labeling potential tasks with difficulty levels to adding concreteness to improve scaffolding. These self-efficacy-related design remedies for supporting gender-diverse populations are also potentially relevant to supporting SES-diverse populations.

Drawing on these design recommendations, we propose the following heuristics for this facet:

Heuristic5: Build confidence by scaffolding unfamiliar tasks.

Heuristic6: Clearly indicate where and how to get help.

### 5.4 Design Recommendations for Facet 4: The Privacy/Security Facet

Two design recommendations pertained to the Privacy and Security facet. Because using technology for social media and communication can raise fears of unwanted information disclosure [78, 120], Guberek et al. recommended wide support for on-demand information hiding, so that sensitive information on individuals' devices will not be released even if their devices are taken from them. Guberek et al. also recommended that technology should communicate potential privacy/security risks upfront, such as by making the flow of the user's information transparent. Woelfer et al.'s research on designing technology for individuals experiencing homelessness suggested that technologies employ separate communication channels that are protected from surveillance [209]. The common theme among these and similar design recommendations are to both increase transparency about how much control users actually have over the privacy/security risks of a particular system and, most critically, to give users more control over their privacy.

Drawing on these design recommendations, we propose the following heuristics:

Heuristic7: Make privacy and security controls transparent and user-controlled.

Heuristic8: Protect against surveillance and support information hiding.

### 5.4 Design Recommendations for Facet 5: The Risk Facet

Two types of design recommendations addressed technology related to the Attitudes toward Risk. Dillahunt et al. researched low-income individuals who live in transportation-scarce environments and use ridesharing services, reporting that not only cost, but also limited payment methods, limited use of smartphones, low digital literacy, lack of trust in the monetary transactions, and variable digital literacy could make these services infeasible. They recommended that ridesharing service providers should coordinate with local businesses to install public kiosks in trusted local businesses such as barbershops and community centers. They proposed that among the advantages of these strategically placed kiosks would be reducing individuals' needs of smartphones, expanding the forms of payment possible, and enabling individuals at the business to pool their digital literacy to help each other navigate the interfaces [57]. In addition, because GenderMag also uses the Risk facet, the design recommendations in the GenderMag Design Catalog that are pertinent to Risk could also be applicable to this facet [75]. Two frequent themes in the GenderMag Design Catalog's risk-related recommendations are a risk-averse user's perceived risk of wasting scarce time on unproductive paths in technology and a risk-averse user's perception of the likelihood of failure. To address barriers like these, many of the Design Catalog's recommendations center on clarifying the steps needed to succeed, which allows users to make informed choices on what to spend their time on, whether their efforts are likely to end in success and what are the ways to recover from failure. One example of this type of recommendation was clarifying the instructions on an open-source software project site, from "...must be installed. Steps may vary in case of MacOS and Windows" to "...a simple set of steps to get your local machine ready... Choose your operating system <has separate links for MacOs, Windows, and Linux concrete instructions>" [75].

Therefore, we propose the following heuristics to accommodate users with diverse SES backgrounds:

Heuristic9: Communicate the cost and risk of features clearly before users take action.

Heuristic10: Make it easy to recover from errors and abandoned paths.

### 5.6 Design Recommendations for Facet 6: Perceived Control/Attitude toward Authority

We located only one design recommendation in SES-related literature pertinent to the Perceived Control/Authority facet. In this design recommendation, Chaudry et al. points out the importance to low-SES individuals of a classic design technique for all users: that every page should incorporate back and home buttons, to enable users to recover from mistaken or dead-end interactions [43]. As many others have also pointed out, this classic technique relates directly to user control. Chaudry et al.'s results suggest that this feature can matter especially to people in lower SES strata. As Insight11 suggests, users who perceive themselves to have little control over technology's outcomes may be particularly likely to accept technology errors, dead-ends, or unfavorable outcomes as outcomes to which they have no recourse. The addition of controls that consistently and explicitly offer alternative pathways out of a problematic technology outcome may help to mitigate Insight11's effect. Similarly, to mitigate the Insight12's, individuals should be provided with immediate value so that they don't require long-term investment to overcome a problem involving technology.

Therefore, we propose the following heuristics:

Heuristic11: Do not treat technology errors as the user's fault.

Heuristic12: Provide immediate value; do not require long-term investment to get started.

## 6 DISCUSSION: ACTIONABILITY

Since our goal for this facet set is actionability, we now consider actionability's intertwinings with validity, utility, and extensibility. For example, knowing whether the facet set is valid determines whether it can be acted upon in good conscience; knowing whether the facet set has utility determines whether acting upon it is likely to be time well spent; and knowing whether it is combinable with other diversity dimensions' facets helps determine whether it can be harnessed to support intersectional identities.

### 6.1 Validity and Utility: Answered Questions and Open Questions

In formally evaluating a facet set like the one presented here, several validity and utility perspectives are possible. The two validity perspectives we consider are correctness (soundness) and completeness.

The answers to these two questions are straightforward: (1) correctness (soundness) is true by construction, and (2) completeness is obviously false (counterexamples in Appendix A). More specifically, soundness asks whether the values of facets presented do actually vary with socioeconomic status. Each facet satisfies this criterion because we selected only facets that satisfied it, by Criterion 2 (Section 3). The literature review summarized in Appendix A details the evidence behind each facet's ties to SES. Completeness is whether the final set of six facets contains all of the SES factors that can impact the way individuals use and experience technology. We know that it does not. For example, Appendix A suggests that there are at least 20 facets potentially useful for a socioeconomic-aware HCI, and our facet set size is only six.

Given that our facet set is a sound *sub*set of SES-diverse characteristics, the next questions are whether this subset is usable and effective. The usability question is largely open. Answering the usability question will require HCI practitioners to start use it in the field, and so far only one such field trial (with positive results) has occurred [2]. As creators of other InclusiveMag-guided facet sets have found, facet sets may need iteration to achieve practical usability. For example, in an academic study of InclusiveMag usage [126], the teams iterated as they attempted to apply their facets to real products. Through their iterations, some teams found that some of their facets were too concrete and needed to be generalized, while others were too general and needed to be more concrete. Likewise, the GenderMag method started with five evidence-based facets, later dropped one as a result of field testing, and still later added a different one as a result of more field testing [35]. As explained in Section 3, Agarwal's field study has already caused a change from five SES facets to six, by splitting off the Privacy/Security facet from the Risk facet [2].

The question of the set's utility in practice also remains largely open. Recall Criterion 6 (Section 3): does the facet set indeed capture a "large enough" subset of SES-diverse characteristics, that designing around them makes an impactful improvement in the SES-inclusivity of the technology? Approaches that have been used to answer these questions for GenderMag facets suggest similar possible steps forward for SES facets. One example is Vorvoreanu et al.'s before/after study on a product redesign based on gender facets [200], and another example is longitudinal testing of the gender facets and the artifacts/methods derived from them (e.g., [84, 139]). Agarwal's field study using SocioeconomicMag [2] produced promising results in this respect, but more research is needed. We encourage other researchers to try out one or more of the actions presented in Section 4, to shed more light on the utility question.

### 6.2 Extensibility to Intersectional Identities

An individual in any SES group also belongs to other identity groups. For example, everyone is in a racial group(s), has a gender identity(ies), and so on. An individual's combined identities are described in the literature as their intersectional identity. In HCI, intersectionality is a framework for engaging with the complexity of users' intersectional identities and situating these identities in their contextual surroundings [63, 153, 168, 210]. Intersectional HCI offers new perspectives on historically excluded populations, such as African Americans and Latinx, who otherwise are studied without accounting for other factors of their identities, such as gender, sexuality, or SES (e.g., [32, 65, 98, 128, 154, 162, 185]).

The fact that the SES facets in this paper were derived according to the InclusiveMag method provides an avenue by which researchers can incorporate the SES facets with other facets to create new, intersectional perspectives. The process for combining InclusiveMag-generated facets from multiple identities to support intersectional identities (e.g., SES facets with age facets [123]) draws from the programming-language-theoretic notion of type abstraction—but abstracting over facet types, *not* people's identities [34]. Unlike reasoning concretely over a finite, potentially inequitable, collection of empirical data that some set of researchers managed to collect, reasoning abstractly over facet types allows *equal representation* of every corner of the infinite set of values possible for every defined facet type and their combinations. Burnett et al. presented a formal proof of the combinatory power of the approach, as well as concrete use-cases and limitations [34].

## 7 CONCLUSION

The digital divide, often framed as a critical aspect of the broader socio-economic divide, is a key problem that our society faces today and is being discussed by researchers, practitioners and policymakers. Still, although many recognize the digital divide in hardware access/quality (e.g., computers, wifi), much less attention has been given to user experiences.

Yet, the results of our synthesis show that the SES-inclusivity of SES-diverse individuals' user experiences needs serious attention. Among these results were:

- *SES-related hardships* in user experiences are far from rare; Section 3 enumerates numerous examples, and more can be found in Appendix A.
- A large fraction of the HCI community's collective body of knowledge about SES-related issues with technology products *can be connected*. The connection "glue" is the set of six concepts we have termed the *SES facets* (Sections 2 and 3): Access to Reliable Tech; Communication (relating to Literacy/Education/Culture); Tech Self-Efficacy; Privacy/Security attitudes; Risk attitudes; and Perceived Control/Attitude toward Authority.
- The SES facets derived from the foundational literature are *actionable*. Sections 4 and 5 identify five actionable byproducts including an expressively powerful vocabulary, a set of design heuristics, multi-personas, a new systematic inclusive design method we call SocioeconomicMag, and a set of design recommendations that map directly to the facets.

Perhaps most important, because the facets are evidence-based, their *validity* (correctness) is given by construction. From this it follows that their byproducts, such as those described above, are also valid, and are ready for HCI practitioners and researchers to use. We also hope that HCI practitioners and researchers can derive even more actionable byproducts, and will help to test and/or improve their *utility* to HCI community members. Ultimately, we hope that these contributions can add to HCI practitioners' practices in ways that help our field ward off SES-inclusivity problems like this one from a low-SES user:

> "It's so scary to see that [enforcement] checks everything about you... My Facebook could … they could probably see it. They could probably hack into my messages... " [78]

## APPENDIX A: THE COMPLETE SET OF FACET CANDIDATES

Table A enumerates the complete literature review we conducted. The number of papers covered for each facet does not reflect the number of papers that exist on that topic but rather the papers we covered via the process explained in Section 2, focusing on SES aspects with ties or potential impacts on HCI. In total, we covered over 200 papers.

The facets and facet examples were defined based on the language of the cited literature and according to what the original authors investigated. For example, if the original authors wrote about race's effect on behavior, we used that paper as foundational literature for race, even if the races written about were minorities in their geographic context. If the original authors instead focused on the minority status, we used that paper for minority status. If the original authors investigated in the senses of both race and minority status, we used that paper as foundational literature for both facets.

Table A: The complete set of facet candidates. All entries in this table satisfied the inclusion criteria for facet candidacy (Section 3).

| | Examples of potential connections to technology usage | References |
|---|---|---|
| Age (groups) | *Example*: Low-SES Children<br>- Most low-SES children wanted to play games, however, were restricted because they either didn't own devices or had to share devices [11].<br>*Example*: Low-SES Teenagers<br>- To encourage low-SES participation in classroom contexts, researchers developed an online tool to allow for reactions to others' assignments [133]<br>*Example*: Low-SES Parents<br>- One goal: to communicate with their family, however, shared devices pose a barrier to coordinating with others. Security issues also a concern with shared devices. [160, 215]<br>*Example*: Low-SES Older Adults<br>- Important to contact remote caregivers to stay connected, but many systems are unreliable [17]. | [11, 13, 15, 18, 30, 46, 66, 96, 102, 116, 133, 144, 148, 149, 152, 160, 161, 164, 204, 208, 215] |
| Authority (Attitude toward) | *Examples*:<br>- Belief in the Just World Hypothesis (see also Control row this table) implies that authority figures are legitimate, leading to a willingness to comply with the authority figures' rules and pronouncements [209]<br>- Low-SES children often raised by parents who tend to defer to authority figures. [184]<br>- High compliance with authority figures' directives and/or perception of low control can prevent negotiation of parameters. For example, low-SES students more likely to turn in incomplete work instead of negotiating with teachers for an extension [39, 87, 147]<br>- Most low-SES individuals don't experiences, practice, or interact with authority figures as their equals, and are also less likely than higher-SES individuals to be critical of authority figures [184]. | [39, 87, 97, 147, 151, 156, 184, 186, 188] |
| Communication Groups | *Example: Online Communities*<br>- Use tech to access online content, but find that the content is not always socially and culturally relevant, especially in developing countries [89]<br>*Example: Families*<br>- Try to expand children's future opportunities (tech literacy) and communicate with family members by using shared devices, owned devices, mobile phones, and older devices, values and technological practice vary from family to family [12]<br>- Low-SES families are more likely to use knock off phones, prepaid phones, older devices, and used devices. For example, rural Kenyan families using low-cost Nokia phones with short battery life spans had to buy new batteries or recharge old ones frequently [212] | [12, 64, 89, 92, 95, 108, 129, 132, 137, 143, 166, 182, 201, 212, 214, 215] |
| Control (Perception of) | *Examples*:<br>- Relatively powerless participants more likely to perceive existing society/economic inequality as fair and legitimate (belief in a "Just World"), perhaps as a way to justify their position in an unequal social system [188]<br>- But in some cases, belief in the Just World hypothesis is positively correlated with perceived control [38, 122, 207]<br>- Poverty is often accompanied by a lack of agency/control [155] | [38, 52, 69, 81, 86, 90, 91, 103, 104, 122, 147, 155, 188, 207, 213] |
| Culture | *Examples*:<br>- Virtual peer tutor helped low- to mid-SES African American students pick up "school English" while also allowing students to remain connected to own dialect and culture [148]<br>- Latino parents have felt excluded from school communities (e.g., due to socioeconomic status, social history, English language mastery). Culturally responsive family-school partnerships may empower minority parents to advocate for their children's education and thereby to place hope in education as a means of social mobility [71] | [7, 45, 71, 73, 80, 127, 136, 148, 157, 202, 195] |
| | - African-American children of low-SES understand narrative information better than children of high SES [73]<br>- Cultural knowledge stores (e.g., household practices, climate adaptations) cultural minority/low-SES groups already have potential resources from which teachers could draw to support students' literacy and content learning, schools with bilingual education programs tend to serve working-class students whose knowledge stores and potential/desire to succeed are often underestimated, further reducing educational quality [127, 195] | |
| Education | *<u>Examples</u>:*<br>- Low-SES individuals tend to receive lower technology-related education than higher-SES individuals [87]. | [1, 10, 13, 25, 46, 49, 60, 72, 87, 96, 135, 148] |
| Employment | *<u>Example</u>: Low-SES Healthcare Professional*<br>- Use wide variety of devices for work and communication but sometimes still lack key technological and financial resources [115]<br>*<u>Examples</u>: Unemployed/underemployed*<br>- Sharing economy has potential to reduce transportation barriers (cost of public and private transport) that make finding sustainable employment difficult for low-SES people [55]<br>- Online job search is attuned to needs of professionals and highly qualified but leaves behind the low-SES unemployed [58] | [55, 58, 115, 121, 181] |
| Empower-ment | *<u>Example</u>: To Protect Community*<br>- Information and communication technologies for grassroots crime prevention empowered 5 low-SES Chicago neighborhoods to protect families and property [64] | [64, 169] |
| Health | *<u>Example</u>: Low-SES Populations with Disabilities*<br>- Assistive tech crucial, but many Low-SES people with hearing, vision, and other disabilities face difficulty using social media or other platforms [41, 191]<br>*<u>Examples</u>: Low-SES Patients*<br>- Low-SES patients with diabetes use individual devices, mobile phones, and desktop computers to self-manage health and communicate with family [22]<br>- Low income disproportionately accompanied not only by ill health but also by difficulties accessing/obtaining health care [155]<br>- Medical resources can be inaccessible/difficult to understand for low income patients [121] | [22, 31, 41, 42, 83, 121, 141, 155, 166, 191, 190] |
| Income | *<u>Examples</u>:*<br>- Technology can be used to access employment/future opportunities through reduced income-related transportation barriers and the idea of a sharing economy [55] | [19, 55, 92, 136, 160, 171, 181, 182, 196, 215] |
| Language, tech communications | Examples:<br>- Some occupations require higher literary engagement than others, with higher-paying occupations tending to require higher literacy engagement. Limited literary engagement (working with menus, receipts, handouts, etc. as opposed to documents and newspapers) is thus a barrier preventing some from higher-paying jobs. For example, many low-income migrants hold jobs with minimal literary engagement, e.g., as field workers, menial employees, unemployed, or part-time [53, 113]<br>- Latino parents have felt excluded from school communities, in some cases due to their English language mastery [71]<br>- Monolingual Spanish-speaking mothers use digital tools like movies, video games, etc. to help raise bilingual children fluent in English but still connected to home culture, thus supporting a multicultural, multilingual education [135]<br>- Technology such as Massively Multiplayer Online Games (MMOGs) can help with second language socialization (players gain fluency in second language by socializing within online community's cultural norms) [170] | [5, 25, 30, 53, 58, 59, 61, 71, 73, 80, 107, 112, 113, 133, 135, 138, 142, 148, 150, 170] |
| Living Situation | Example: People experiencing homelessness<br>- Use technology to stay safe and in touch with loved ones. People without homes often use public devices and older devices due to not owning device themselves, which provide limited/unreliable access to tech [105, 209]<br>Example: Low-SES Location<br>- Living or working in dense/unsanitary conditions can prevent following standard health guidelines (like COVID-19 social distancing); reasons can include close proximity to family/co-workers and employers' lack of concern for workers' health [4, 145] | [4, 48, 55, 57, 62, 74, 101, 105, 109, 134, 137, 145<br>146, 159, 166, 175, 178, 183, 202, 209, 211] |
| Minority Status | Example: Immigrants<br>- Immigrants (ethnic, national minority) use tech to communicate with friends, family, community, & institutions to create a collective identity and protect each other with information on immigration enforcement, etc. [78]<br>- Minority racial groups in the U.S. are more likely to experience poverty than Caucasians [155] | [3, 6, 7, 17, 37, 45, 51, 60, 68, 78, 80, 93, 99, 107, 133, 155, 168, 171, 177, 215] |
| Planning (Willingness to) | Examples:<br>- Individuals' propensities to plan ahead vary by socioeconomic status [24, 54, 125]<br>- Racial and ethnic differences within Low-SES populations varied in terms of advance care planning [40] | [24, 40, 54, 125] |
| Privacy / Security | Examples:<br>- Asked about concerns and perceived risks of technology use, many undocumented immigrants mentioned security-related threats that are not unique to their immigration status— concerns such as identity theft, online financial fraud, unauthorized access to their Facebook accounts, and hackers who might steal their information or impersonate them online [78]<br>- The participating undocumented immigrants primarily managed privacy and security concerns through three practices: network regulation, varying degrees of self-censorship, and deliberate choice of communication channels [78].<br>- Another privacy practice by some low-SES individuals: deleting content; e.g., deleting entire threads or histories of content, specific chats, specific media, or specific queries [165]<br>- The expectation that individuals should manage their own privacy is unrealistic and problematic for vulnerable populations [189] | [18, 29, 78, 120, 124, 165, 180, 189, 196] |
| Race | *Examples:*<br>- On intersectionality: race is closely intertwined with SES, with racial minorities disproportionately more likely to be in low-SES situations (low income, low education, poor healthcare, etc.) [50]<br>- Caucasians more likely to use computers and to have higher rates of internet access and broadband internet than African-Americans, who are more likely to use mobile phones and play/buy video games [215] | [16, 35, 48, 50, 215] |
| Risk (Attitude toward) | *Examples:*<br>- Low-SES individuals are exposed to greater technological risk than high-SES people. For example, Latinx immigrants are at risk of online bullying/harrassment [78]<br>Risk tolerance varies by socioeconomic status [77, 124] | [10, 77, 78, 120, 124, 196] |
| Self-efficacy | *Examples:*<br>- Self-efficacy tends to be a self-fulfilling prophecy [21].<br>- Access to technology (see Tech Situation row) is one determining factor for computer self-efficacy [190]<br>- Technology experience increases technology self-efficacy but low-SES individuals have fewer opportunities than higher-SES individuals to gain experience with technology due to their reliance on shared devices [215] | [1, 82, 111, 167, 187, 194, 193, 215] |
| Tech situation | *Example: Tech Literacy:*<br>- Distrust in technology did not deter low-SES populations' willingness to use technology [99]<br>*Examples: Trust in Technology:*<br>- Latino migrant minorities had security concerns in technology [78]<br>- Some low-SES individuals felt that technology was insecure, untrustworthy, and subject to attack [196]<br>*Example: Tech* Access*:*<br>- Low-SES youth were more likely than the adults to use technology, so they often relayed information to adults through shared devices [68]<br>- 62% of households living in poverty have computers, and of those, fewer than half have internet access [67].<br>- Some low-SES individuals depend on mobile data for access to the internet. Convenience and ease-of-use were reasons for this mobile-primary lifestyle [78]<br>- Low-SES students' schools less likely to offer computer-related courses [117, 119, 203] | [27, 48, 57, 67, 68, 78, 99, 101, 105, 117, 118, 119, 142, 158, 173, 174, 192, 193, 196, 202, 203, 209, 215, 47] |
| Time/ Availability | *Example:*<br>- Low-SES individuals often have to work multiple jobs, which impacts their time availability [8, 13]<br>- Low-SES families spend more time and energy than higher-SES families feeding, clothing, and caring for their children. Their time also goes to complicated logistics due to use of public transportation and reliance on bureaucracies for income, food stamps, etc. [216] | [8, 14, 44, 216] |

## APPENDIX B: THREE SES MULTI-PERSONAS

Dav (David, Davu, Davida)

**Introduction:**

**Background/interests:**

***Access to Reliable Technology:*** Dav has spotty access to reliable devices with reliable internet access, so relies mainly on a mobile phone for internet access. Dav also often uses shared devices or public devices to get work done. This affects how, when, and why Dav uses technology.

***Communication Literacy/Education/Culture:*** Dav went to school in a low-SES community which offered only a basic education. Now Dav rarely chooses to read lengthy or complex text (e.g., newspapers), and some cultural/literary allusions are unfamiliar to Dav. Although the school had a few older computers, it offered little technology education.

***Attitudes toward Technology Risks:*** Dav is risk-averse. For example, Dav's life is crowded, so they rarely have spare time. So Dav is risk-averse, such as about using unfamiliar technologies that they might need to spend extra time on, even if the new features might be relevant. Dav instead performs tasks using familiar features, because they're more predictable about what Dav will get from them and how much time they will take.

***Technology Privacy and Security:*** Dav is very protective of their personal information, like their location and identity. Dav's caution stems from their privacy/security being particularly at risk because of having to share devices, prior negative experiences with high surveillance, prior experiences with credit card/identity theft, etc.

***Perceived Control and Attitude toward Authority:*** Dav does not expect to have much influence over technology's outcomes. Instead, Dav views technology as if it represents an authority figure, so expects technology to treat Dav as other authority figures do.

***Technology Self-Efficacy:*** Dav's prior experiences and education have produced a lower technology self-efficacy than their peers about using unfamiliar technology features. If problems arise with technology, Dav often blames themselves for these problems. This affects whether and how they will persevere with a task.

Figure B-1: "Dav", the customizable multi-persona whose facet values are at the lower-SES endpoints. (Also shown in the main paper.)

Ash (Asha/Ashwin)

**Introduction:**

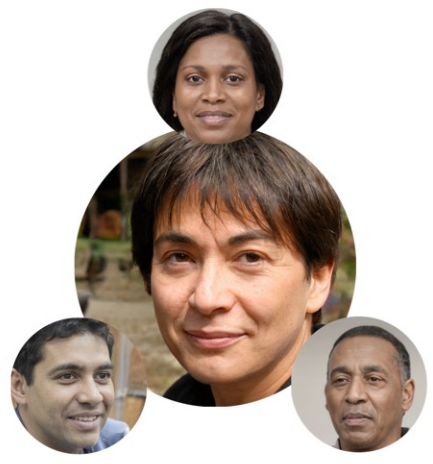

**Background/interests:**

***Access to Reliable Technology:*** Ash has fairly consistent access to reliable devices and to reliable internet. Ash relies on their own devices and rarely uses a shared device or a public device to get work done. This affects how and when Ash uses technology.

***Communication: Literacy/Education/Culture:*** Ash had access to average-quality education growing up and was exposed some technologies. Due to their education, they have some experience and don't struggle much with communications using implicit assumptions, cultural references, jargon, or complex sentence structures.

***Attitudes toward Technology Risks:*** Ash is risk-averse. For example, they're risk-averse about using unfamiliar technologies that they might need to spend extra time on, even if the new features might be relevant. Ash instead performs tasks using familiar features, because they're more predictable about what Ash will get from them and how much time they will take.

***Technology Privacy and Security:*** Ash has never lost personal information, like their identity, but worries some about people tracking their location. Still, because they tend to rely on their own device, their perception of technological features as being risky is medium due to their past and familiarity with technology.

***Perceived Control and Attitude toward Authority:*** Ash views technology's output as a suggestion that they can challenge, change, or work around. As a result, Ash often feels that they have control in the experiences they have when they use technology.

***Technology Self-Efficacy:*** Ash has average technology self-efficacy, compared to their peers, about doing unfamiliar computing tasks. And if problems do arise with technology, Ash might blame the technology for the problems, or might blame themself. This affects how long they will persevere with a task.

Figure B-2: "Ash", the customizable multi-persona who has a mix of facet values.

Fee (Feechi/Felienne/Felix)

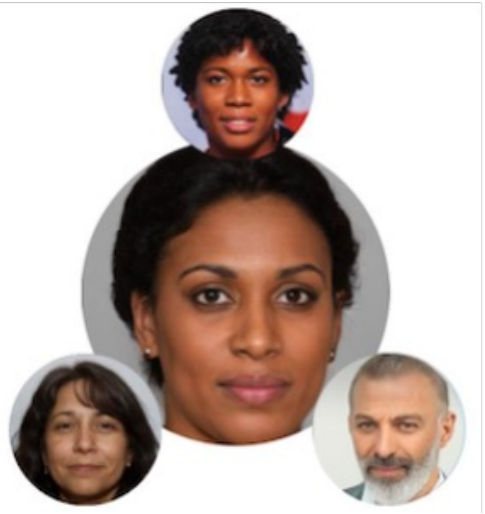

**Introduction:**

**Background/interests:**

***Access to Reliable Technology:*** Fee has high access to reliable devices and to reliable internet. Fee relies on their own devices and rarely uses a shared device or a public device to get work done. This affects how and when Fee uses technology.

***Communication Literacy/Education/Culture:*** Fee had access to high quality education growing up and was exposed to a variety of technologies. They also have more experience, so struggle less with software that uses implicit assumptions, cultural references, jargon, or complex sentence structures

***Attitudes toward Technology Risks:*** Fee doesn't mind taking risks using features of technology that haven't been proven to work. When Fee is presented with challenges because they have tried a new way that doesn't work, it doesn't change their attitudes toward technology.

***Technology Privacy and Security:*** Fee has never lost personal information, like their identity, and is generally not too worried about people knowing their location. Because they tend to rely on their own device, their perception of technological privacy/security risks is low due to their past and familiarity with technology.

***Perceived Control and Attitude toward Authority:*** Fee views technology's output as a suggestion that they can challenge, change, or work around. As a result, Fee often feels that they have control in the experiences they have when they use technology.

***Technology Self-Efficacy:*** Fee has higher technology self-efficacy than their peers about doing unfamiliar computing tasks. And if problems do arise with technology, Fee blames the technology for the problems, not themself. This affects whether and how they will persevere with a task.

Figure B-3: “Fee”, the customizable multi-persona whose facet values are at the higher-SES endpoints.